\documentclass[prl,twocolumn,amssymb]{revtex4}

\newcommand{\w}{{\omega}}

\def\be{\begin{eqnarray}}
\def\ee{\end{eqnarray}}
\newcommand{\nn}{\nonumber\\}

\newcommand{\Eq}[1]{Eq.~(\ref{#1})}

\newcommand{\p}{\partial}
\newcommand{\ua}{\uparrow}
\newcommand{\da}{\downarrow}

\usepackage{graphicx}

\begin{document}

\title{Effects of a Fluctuating Interface between a Superfluid and a Polarized Fermi Gas}
\author{Hui Zhai$^{1,2,3}$ and Dung-Hai Lee$^{1,2}$}
\affiliation {1. Department of Physics, University of California,
Berkeley, California, 94720, USA\\ 2. Materials Sciences Division,
Lawrence Berkeley National Laboratory, Berkeley, California, 94720,
USA\\
3. Center for Advanced Study, Tsinghua University, Beijing, 100084,
China}
\date{\today}
\begin{abstract}
Motivated by recent experiments in trapped Fermi gas with spin
population imbalance, we discuss the effects of the quantum and
thermal fluctuations of the interface between a fully paired
superfluid core and a fully polarized Fermi gas. We demonstrate that even if there is no true partially polarized thermodynamic phase in bulk, the interface fluctuation
can give rise to a partially polarized transition regime in trap.
Our theory yields a definite prediction for the functional forms of
the spatial profile of spin polarization and pairing gap, and we show
that the spin-resolved density profiles measured by both the
MIT and Rice groups obey this function form. We also show that sufficient
large fluctuation will lead to a visibly unequal density even at the center of the cloud. We hope this picture can shed lights on the controversial
discrepancies in recent experiments.

\end{abstract}
\maketitle

Phase separation is an ubiquitous natural phenomenon. Its root is
first order phase transition. For example, as a function of
temperature ($T$) and magnetic field ($H$), an easy-axis ferromagnet
phase separates into two oppositely magnetized phases when $(T,H)$
are {\it fined tuned} to the boundary of the first-order transition
($H=0, T=T_c$). On the other hand, as long as $T<T_c$ phase
separation occurs without needing fine tuning when the
magnetization, rather than magnetic field, is fixed. In the presence
of phase separation the dynamics of the interface determines the low
energy/temperature physical properties of the system.
Recently experiments of ultracold Fermi gases near Feshbach resonance
once again reveal phase separation
phenomenon\cite{early,mit-ps,rice-ps}. In these experiments, an
external-imposed spin population imbalance frustrates the strong
pairing interaction due to Feshbach resonance. As a result, the
system phase separates into a core of superfluid with equal spin
population and a shell of excess unpaired fermions.

Although similar
phenomena were observed, there are important difference between the
findings of two major experimental groups at MIT and Rice
University. Let $N_{\uparrow,\downarrow}$ represent the number
of spin up/down atoms, and
$P=(N_{\uparrow}-N_{\downarrow})/(N_{\uparrow}+N_{\downarrow})$ be
the population imbalance. Here are the highlight of the differences.
The MIT group finds (i) for small $P$, the spin-resolved density
profile exhibits three regions: a superfluid core with
$n_{\uparrow}=n_{\downarrow}$, a partially polarized (PP) shell with
$n_{\uparrow}>n_{\downarrow}>0$ and a fully polarized region with
$n_{\downarrow}=0$. The width of the PP shell is as
wide as the other two regions (see Fig.5(a) of Ref.\cite{mit-ps}).
(ii) The interfaces between different regions always follow the
equal potential contour as predicted by theories using local density
approximation. (iii) As $P$ exceeds a critical value
$P_{\text{c}}\sim 70\%$, the equal density core disappears and also
the condensation fraction vanishes, which is referred to as the
Chandrasckhar-Clogston (CC) limit of superfluidity.\cite{mit-ps} In
contrast, the Rice group finds (i) the width of the PP region is much narrower than the width of other two
regions (see Fig.3(b) of Ref.(\cite{rice-ps})). (ii) At large $P$
the interface does not follow the equal potential contour. (iii) The
equal density core exists up to the highest $P$ studied ($\sim
95\%$), and there is no sign of CC limit.\cite{rice-ps}
One can expect the explanation of these discrepancies to shed light
on a number of fundamental issues of this problem. For examples, what
is the nature of the PP region, does it reveal a
true quantum phase of the system, and what controls its width; does
the CC limit for superfluidity exists, and if so, what is the
mechanism.

\begin{figure}[bp]
\begin{center}
\includegraphics[angle=0,scale=0.5]
{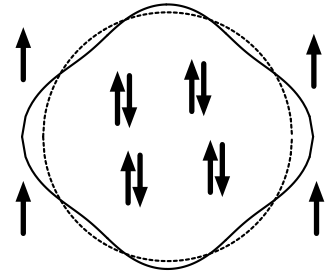}\caption{Schematic of the interface flucuation. The
dashing line is the static location of the interface which separates
a fully paired superfluid phase and a fully polarized fermi gas. The
solid line sketches an example of surface mode which distorts the
interface.\label{illu-surface}}
\end{center}
\end{figure}

These experiments have stimulated much theoretical discussions in
the literature. Most of the existing theories  assume the system to
follow the confining potential {\it adiabatically}, so that locally
it corresponds to a bulk phase with the chemical potential
$\mu_{\ua,\da}=\mu_{0\ua,\da}-V({\bf r})$, where $\mu_{0\ua,\da}$ is
a spin-dependent constant and $V({\bf r})$ is the trapping
potential. The issue at hand is whether the PP state is a stable
bulk phase near Feshbach resonance. By comparing mean-field
energies, many authors predicted a narrow window of PP phase in the
bulk phase diagram\cite{Leo}. However due to the approximate nature
of the treatment, and the fact that even in mean-field theory the
energy difference between different phases is tiny, it is hard to
know whether such a PP bulk phase really exists or not.

In the following we make a bold assumption: the PP region found in experiments is {\it
not} a true phase, rather it is due to the fluctuations of the
interface between the equal-population superfluid and the fully
polarized phase. Our assumption implies that, for interaction
strength relevant to the experimental systems, there is a direct
zero-temperature first order transition between the superfluid and the fully
polarized phases as a function of the Zeeman field. Our proposal
is motivated by two recent developments. (i) Recently the MIT group
found the presence of a full pairing gap in the PP state at $P>P_c$
where the superfluidity is destroyed\cite{Rf}. This contradicts
theories which attribute the CC limit to
depairing\cite{CClimit}. (ii) The importance of interface physics is
highlighted by recent theoretical works which attribute the
deformation of the interface from the equal potential surface to the
existence of finite surface tension\cite{Erich,Stoof}. Of course,
interface physics involves not only statics but also dynamics .

In the rest of the paper we analyze the effect of {\it harmonic}
interface fluctuations between the superfluid and the fully
polarized phase. In Fig.\ref{illu-surface} we illustrate a snap shot
of an allowed interface deformation where the enclosed volume is
kept fixed. Surface fluctuations have been studied in early days of
atomic BEC and fermion superfluid with equal population\cite{review}. However
since the system is consisted of only one single phase, such
fluctuation is less important as it only affects the low density
region at the edge of the cloud. In comparison the effects of
interface fluctuation are far more pronounced here which lead to a very
precise and universal profile of spin polarization and
superfluid order parameter
across the two phases. 

Here are how we describe the interface. i) Motivated by the
experimental facts\cite{rice-ps,mit-ps}, we assume the equilibrium
interface ${\cal S}$ is a closed two-dimensional surface with
cylindrical symmetry: \be {\cal S}=\{(x,y,z): x^2+y^2+\epsilon^2(z)
z^2=a^2\}.\label{interf}\ee Here $\epsilon(z)$ is an even function,
its value is unity for $z=0$, and decreases monotonically to
$a/z_0<1$ as $z$ approaches $\pm z_0$, the north and south poles of
${\cal S}$. In the following we shall use $z$ and the azimuthal
angle $\phi$ as the coordinates for ${\cal S}$, so that
$x=\sqrt{a^2-\epsilon^2(z)z^2}\cos\phi, ~
y=\sqrt{a^2-\epsilon^2(z)z^2}\sin\phi,~ z=z$. It should be apparent
that this is a locally orthogonal coordination, i.e.,
$ds^2=g_{\phi}(z)d\phi^2+g_{z}(z)dz^2.$  (ii) We describe the
fluctuation of the interface from its equilibrium shape by a {\it
harmonic} quantum elastic model. Let $u(z,\phi,t)$ be the
time-dependent {\it normal displacement} of the interface from its
equilibrium position, the Lagrangian of our model reads \be
\mathcal{L}=\int_{\cal S} d\phi dz \sqrt{g_\phi g_z}\Big[{m\over
2}\dot{u}^2-{V(z)\over 2}u^2-\sum_{\mu=\phi,z}{(\p_\mu u)^2\over
2g_\mu}\Big],\label{La} \ee where $m$ is an introduced phenomenological parameter. The normal modes of the interface
fluctuation obey the following differential equation \be
\frac{1}{\sqrt{g_\phi g_z}}\p_z\left(\sqrt{{g_\phi\over g_z}}~\p_z
u_\alpha\right)+{~\p_\phi^2 u_\alpha \over g_\phi}
-V(z)u_\alpha=-m\w_\alpha^2 u_\alpha,\label{eig}\ee where
$\w_\alpha$ is the normal mode frequency. In later analysis we shall
expand $u(z,\phi,t)$ in terms of the normal modes :\be
&&u(z,\phi,t)={\sum_\alpha}^\prime A_\alpha(t)
u_\alpha(z,\phi).\label{exp}\ee Note that all quantities in the
above equation are real. In \Eq{exp} the sum excludes the
``breathing'' mode, the mode whose associated distortion does not
conserve the enclosed volume. Interestingly, despite the presence of
$g_{\phi}(z)$ and $g_{z}(z)$ in \Eq{La} and \Eq{eig}, the functional
form of the interface profile can be determined exactly along
special directions such as the $z$-axis\cite{rice-ps} and any radial
direction in the $z=0$ plane\cite{mit-ps}.

\begin{figure}[bp]
\begin{center}
\includegraphics[angle=0,scale=0.5]
{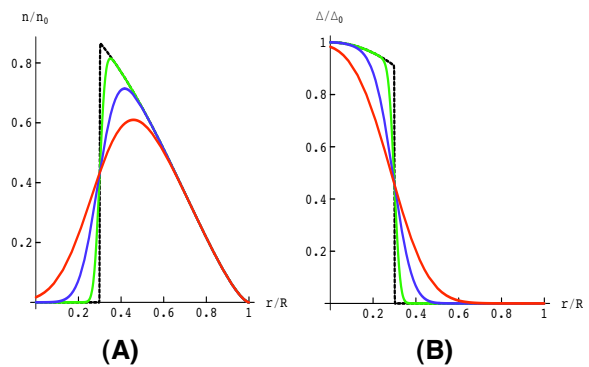}\caption{Plots of \Eq{prof} and \Eq{n0} for different
choices of $\xi/R$. We have use $r_0/R=0.3$ for all curves.
$\xi/R=0.03$ for green line, $\xi/R=0.1$ for blue line and
$\xi/R=0.2$ for red line. The black dashing lines are the profiles
without interface fluctuation, which exhibit discontinuity.
\label{profile}}
\end{center}
\end{figure}

For example, as the main result of this paper, we obtain the
following function form for the spin polarization
$M=n_{\uparrow}-n_{\downarrow}$ and the pairing gap profiles along a
high symmetry radial direction ($\hat{t}$): \be
M(r,\hat{t})&=&\frac{1}{2}\left[1+\text{Erf}\left(\frac{r-r_0(\hat{t})}
{\xi(\hat{t})}\right)\right]n_{\rm{p}}(r,\hat{t})\nn
\Delta(r,\hat{t})&=&\frac{1}{2}\left[1-\text{Erf}\left(\frac{r-r_0(\hat{t})}
{\xi(\hat{t})}\right)\right]\Delta_{{\rm s}}(r,\hat{t}).
\label{prof} \ee  Here Erf($x$) is the error function,
$n_p(r,\hat{t})$ is the density profile of a fully polarized Fermi
gas, and $\Delta_{\text{s}}(r,\hat{t})$ is the pairing gap profile
of the fully-paired superfluid, along the same direction in the same
trap. In particular we use\cite{rf-insitu} \be
&&n_{\text{p}}(r,\hat{t})=n_{0}[~1-(r/R(\hat{t}))^2~]^{3/2}\nn
&&\Delta_{\rm
s}(r,\hat{t})=\Delta_{0}[~1-(r/R(\hat{t}))^2~].\label{n0}\ee In
writing down the first line of \Eq{n0} we have assumed the
temperature is sufficiently low and the trapping potential is harmonic. For the
second line of \Eq{n0} we have taken into account the universality
of the pairing gap near Feshbach resonance\cite{universality}.

\begin{figure}[tbp]
\begin{center}
\includegraphics[angle=0,scale=0.38]
{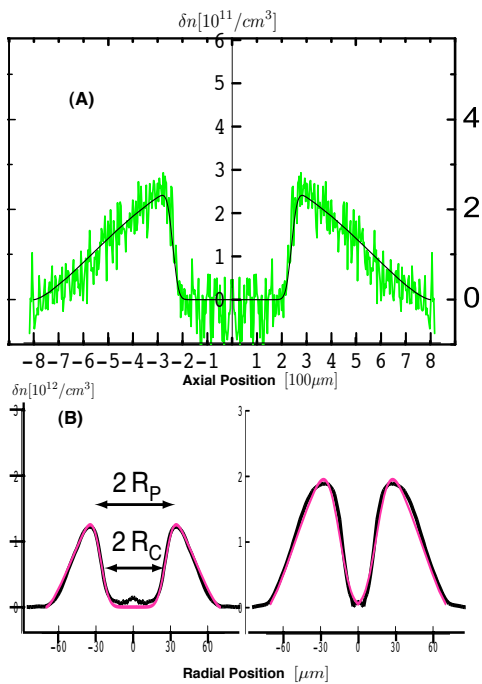}\caption{Fit the measured spin polarization profile to
\Eq{prof} and \Eq{n0}. The green line in (A) and the black line in
(B) are the data of the Rice\cite{rice-ps} and MIT\cite{mit-ps}
group, respectively. The black line in (A) and the purple line in
(B) are our best fits. The fitting parameters for (A) are
$r_0=240\mu m$, $\xi=80\mu m$, $n_0=2.97\times
10^{11}\text{cm}^{-3}$ and $R=800\mu m$. The fitting parameters for
the left and right panels of (B) are $r_0=27\mu m$, $\xi=23\mu m$,
$n_0=2.058\times 10^{12} \text{cm}^{-3}$ $R=70\mu m$, and $r_0=16\mu
m$, $\xi=35\mu m$, $n_0=2.645\times 10^{12}\text{cm}^{-3}$, $R=73\mu
m$, respectively.\label{fitting}}
\end{center}
\end{figure}

An appealing feature of \Eq{prof} is that the microscopic
information about the interface (and temperature) enters through a
single parameter $\xi(\hat{t})$. In Fig.(\ref{profile}) we plot
\Eq{prof} and \Eq{n0} for typical values of parameters. For $r_0\gg
\xi$, the density difference at the center of the cloud is
exponentially small which reveals an equal density core. For small
$\xi$ (green line), one can have rather narrow intermediate
partially polarized regime (like the Rice group's result). Wider
partially polarized regime (like the MIT group's result) can
be obtained when $\xi$ is given a larger value (blue line). When
$\xi$ increases toward $r_0$, hence the interface fluctuation becomes
more severe, even the center of the core can become visibly
partially polarized. Meanwhile the pairing gap $\Delta_0$ remains
finite similar to the result of Ref.\cite{Rf}. Clearly in this
theory the partially polarized region is caused by the interface
fluctuation rather than a PP normal phase with a vanishing
(or smaller) pairing gap.

In Fig.(\ref{fitting}) we fit the polarization profile measured by
both the MIT and Rice group to the first line of \Eq{prof} and
\Eq{n0}. These fits are excellent! Recently a new scheme of
tomographic rf spectroscopy invented in MIT\cite{rf-insitu} gives
the promise of measuring the spatial profile of pairing gap
$\Delta(r)$, which can be used to test the second line of
Eq.(\ref{prof}). For the same sample, if one measures both $\delta
n(r)$ and $\Delta(r)$, and provided that both $\delta n(r)$ and
$\Delta(r)$ can be well fit by \Eq{prof}, one can extract the values of $r_0$
and $\xi$ by fitting these two spatial profiles to these two
functions. A falsifiable predication of our phenomenological theory
is that the values of $r_0$ and $\xi$ extracted from these two
spatial profiles, $\delta n(r)$ and $\Delta(r)$, shall be the same.
This is because in both case it is the same interface
fluctuation that gives rise to the smooth spatial profiles.

Now we fill in the details of how to obtain \Eq{prof} from the model
defined in \Eq{La} and \Eq{eig}. We will first consider the effects
of quantum fluctuation at $T=0$. Later we shall generalize the
result to finite temperature.

Substituting \Eq{exp} into \Eq{La} and use the orthonormal relation
between different normal modes \be \int_{\cal S}\sqrt{g_\phi g_z}
d\phi dz u_\alpha(z,\phi)u_\beta(z,\phi)=\delta_{\alpha,\beta}\ee we
obtain
\begin{eqnarray}
\mathcal{L}={m\over 2}{\sum_\alpha}^\prime \left[
\dot{A}_\alpha^2-\w_\alpha^2 A_\alpha^2\right]. \label{harmonics}
\end{eqnarray}
\Eq{harmonics} describes a set of independent harmonic oscillators.
The corresponding Hamiltonian is given by \be
\mathcal{H}={\sum_\alpha}^\prime \left[ {1\over
2m}\Pi_\alpha^2+{m\w_\alpha^2\over 2}
 A_\alpha^2\right],
\label{ham} \ee where $\Pi_\alpha$ and $A_\beta$ obey the following
commutation relation
$[A_\beta,\Pi_\alpha]=i\hbar\delta_{\alpha\beta}.$ The ground state
wavefunction of \Eq{ham} is given by
\begin{eqnarray}
\Psi\left[\{A_\alpha\}\right]={\prod\limits_{\alpha}}^\prime
\left({m\w_{\alpha}\over
\hbar \pi}\right)^{1/4}~\rm{Exp}\left[-{m\w_{\alpha} \over 2\hbar }A_\alpha^2\right],\label{wave-function}
\end{eqnarray}
where $\prod^\prime$ excludes the breathing mode from the product.
Here we explain why some directions ($\hat{t}$ in \Eq{prof} and
\Eq{n0}) are special. For a generic point on the surface specified
in \Eq{interf}, the direction of surface normal does not agree with
the vector connecting the center to the point in question. There are
special points on the surface, for which these two directions agree.
For the surface described by \Eq{interf} these special directions
are $\pm\hat{z}$ and $\hat{r}$ around the equator at $z=0$. Let
$(z_0,\phi_0)$ be such a special point, \be
u(z_0,\phi_0)={\sum_\alpha}^\prime A_\alpha
u_\alpha(z_0,\phi_0).\label{u}\ee Let us first calculate the
probability density for the interface to move by amount $\eta$
\begin{eqnarray}
P(\eta)=\int{\prod\limits_{\alpha}}^\prime
dA_\alpha\delta(\eta-u(z_0,\phi_0))|\Psi[\{A_\alpha\}]|^2 \label{Pd}
\end{eqnarray}
By substituting \Eq{u} for $u(z_0,\phi_0)$, and making use of the identity that
\begin{eqnarray}
\delta(\eta-u(z_0,\phi_0))=\int {d\lambda\over 2\pi}
e^{-i\lambda[\eta-u(z_0,\phi_0)]},\label{trick}
\end{eqnarray}
one can then integrate out each $A_\alpha$ separately and yield \be
P(\eta)&=&\int_{-\infty}^\infty {d\lambda\over 2\pi}
e^{-i\lambda\eta} e^{-{{\lambda^2\over 4}\sum_\alpha}^\prime
m\w_\alpha
u^2_\alpha(z_0,\phi_0)/\hbar}\nn&=&{1\over\sqrt{\pi\xi^2}}e^{-{\eta^2/\xi^2}}.\label{probabilityd}
\ee In the above \be \xi^2={m\over \hbar}{\sum_\alpha}^\prime\w_\alpha
u^2_\alpha(z_0,\phi_0).\ee For a point $r$ away from the center of the
superfluid core, along the vector connecting the center to
$(z_0,\phi_0)$ on the interface, we can calculate the probability
$W(r)$ for it to be in the polarized fermi gas region:
\begin{equation}
W(r)=\int_{-\infty}^{r-r_0}
P(r^\prime)dr^\prime=\frac{1}{2}\left[1+\text{Erf}\left(\frac{r-r_0}{\xi}\right)\right],\label{Pb}
\end{equation}
Therefore the averaged spin polarization is $W(r)n_{\text{p}}(r)$
and the average pairing gap is $(1-W(r))\Delta_{\text{s}}(r)$, i.e.
the results of \Eq{prof}.

Now we consider non-zero temperature. In that case we replace
\Eq{Pd} by \be P(\eta)=\Big <\delta(\eta-{\sum_\alpha}^\prime
A_\alpha u_\alpha(z_0,\phi_0))\Big
>,\label{Pd1}\ee where $<...>$ denotes the quantum statistical
mechanical average. A convenient way to calculate the average of
\Eq{Pd1} is to perform Feynman's path integral in imaginary time.
Let $A_\alpha(\tau)$ be the Feynman path of each harmonic oscillator
in \Eq{ham}, we have \be P(\eta)={\int
{\prod\limits_{\alpha}}^\prime {\cal
D}A_\alpha(\tau)~\delta(\eta-{\sum_\alpha}^\prime A_\alpha(0)
u_\alpha(z_0,\phi_0))e^{-S}\over {\cal Z}}.~ \label{Pd2}\ee Here
${\cal Z}$ is the partition function, and the action $S$ is given by
\be S={m\over 2}\int_0^\beta d\tau {\sum_\alpha}^\prime[
\dot{A}_\alpha(\tau)^2+\w_\alpha^2 A_\alpha(\tau)^2].\ee After using
\Eq{trick} to represent the delta function, we are left with a
quadratic path integral which can be performed exactly to give \be
P(\eta)=\sqrt{\frac{1}{\tilde{\xi}^2\pi}}e^{-\eta^2/\tilde{\xi}^2},
\ee where \be \tilde{\xi}^2={2\beta\over
m}{\sum_\alpha}^\prime\sum_n {u^2_\alpha(z_0,\alpha_0)\over
(\w_n^2+\w_\alpha^2)},\ee and $\w_n$ is the bosonic Matsubara
frequency. As $P(\eta)$ has the same form as
Eq.(\ref{probabilityd}), $W(r)$ will also have the same form as
Eq.(\ref{Pd}) which leads to the same function form of \Eq{prof}.

In conclusion, we have demonstrated that even if there is no true PP
phase in bulk, a PP regime can emerge from the quantum and/or
thermal fluctuation of the interface. The width of PP regime depends
on the strength of the surface fluctuation, and sufficient strong
fluctuation can even lead to disappearance of the equal density
core, with the pairing gap remaining finite. Though the microscopic
description of interface fluctuation remains to be explored, we have
shown that, as long as the theory for surface motion is quadratic,
there is a universal function form for spin polarization and pairing
gap, which can be compared with existing experiments and tested by
future experiments as a justification of this scenario.

{\it Acknowledgment.} HZ thanks KITP of UCSB (supported by the NSF
under Grant No. PHY05-51164), Institute Henri
Poincar\'e workshop on ``Quantum Gases" and KITPC workshop on ``Quantum Phases of Matter" for hospitality, he also would like to thank Erich
Muller for helpful discussions. DHL was supported by the
Directior, Office of Science, Office of Basic Energy Sciences,
Materials Sciences and Engineering Division, of the U.S. Department
of Energy under Contract No. DE-AC02-05CH11231.

\end{document}